\documentclass[12pt,preprint]{aastex}

\shorttitle{Meridional flux transport speed in the photosphere} 
\shortauthors{\v{S}vanda et al.}

\begin{document}

\title{Speed of Meridional Flows and Magnetic Flux Transport on the Sun}


\author{Michal \v{S}vanda, \altaffilmark{1}$^{,}$\altaffilmark{2},
Alexander G. Kosovichev\altaffilmark{3}, and Junwei Zhao\altaffilmark{3}}

\altaffiltext{1}{Astronomical Institute (v.v.i.), Academy of Sciences, Ond\v{r}ejov Observatory, CZ-25165, Czech Republic}
\altaffiltext{2}{Astronomical Institute, Charles University, Prague, CZ-18200, Czech Republic}
\altaffiltext{3}{W. W. Hansen Experimental Physics Laboratory, Stanford University, Stanford, CA~94305-4085, USA}
\altaffiltext{*}{Emails: michal@astronomie.cz, sasha@sun.stanford.edu, junwei@sun.stanford.edu}

\begin{abstract}
We use the magnetic butterfly diagram to determine the speed of the magnetic flux
transport on the solar surface towards the poles. The manifestation of the flux transport is clearly visible as
elongated structures extended from the sunspot belt to the polar
regions. The slopes of these structures are measured and
interpreted as meridional magnetic flux transport speed. Comparison with the
time-distance helioseismology measurements of the mean speed of the meridional flows at the depth of 3.5--12~Mm shows a
generally good agreement, but the speeds of the flux transport and the meridional flow
are significantly different in areas occupied by the magnetic field.
The local circulation flows around active regions, especially the strong equatorward flows
on the equatorial side of active regions affect the mean velocity profile derived
by helioseismology, but do not influence the magnetic flux transport.
The results show that the mean longitudinally averaged meridional flow measurements by helioseismology
may not be used directly in solar dynamo models for describing the magnetic flux transport, and that
it is necessary to take into account the longitudinal structure of these flows.

\end{abstract}


\section{Introduction}

The largest-scale velocity fields on the Sun consist of differential
rotation and meridional circulation. The differential rotation is
defined as an integral of the zonal (East--West) component of the velocity field
depending on the solar latitude, $b$, and radius. The meridional flow 
is calculated as an integral of the North--South component of
the velocity field, generally depending again on the latitude and radius.
Both the differential rotation and meridional circulation are the key
ingredients of the solar dynamo. The differential rotation plays an
important role in generating and strengthening of toroidal magnetic
field inside the Sun, while the meridional flow
transports the magnetic flux towards the solar poles resulting in
cyclic polar field reversals \citep[for a recent review,
see][]{Brandenburg2005}.

The meridional flux transport seems to be an essential agent
influencing the length, strength and other properties of solar
magnetic cycles. Generally, the slower the meridional flows are, the
longer the next magnetic cycle is expected. Dynamo models showed
that the turn-around time of the meridional cell is between 17 and 21
years, and that the global dynamo may have some kind of memory lasting
longer than one cycle \citep{2006GeoRL..3305102D}.

The speed of the meridional flow and its variation with the solar cycle measured by local helioseismology
in the subsurface layers of the Sun were used
as an input in the recent flux-transport models \citep{2006ApJ...649..498D}. In local helioseismology measurements
(e.g. \citeauthor{2004ApJ...603..776Z}
\citeyear{2004ApJ...603..776Z}, \citeauthor{2006ApJ...638..576G} \citeyear{2006ApJ...638..576G}), the meridional flow was
derived from a general subsurface flow field by averaging the
North-South component of the plasma velocity over longitude for a Carrington
rotation period. The studies revealed that the mean meridional flow
varied with the solar activity cycle. These variations may
significantly affect solar-cycle predictions based on the solar dynamo models, which assume that the
magnetic flux is transported with the mean meridional flow speed
\citep{2006ApJ...649..498D}.

Our goal is to verify this assumption and to investigate the
relationship between the subsurface meridional flows and the flux
transport. In this study, we show that the mean meridional flows
derived from the time-distance helioseismology subsurface flow maps
are affected
by strong local flows around active regions in the activity belts. 
These local flows have much less significant effect on the magnetic flux
transport.

\section{Method of measurements}

The magnetic field data were obtained from Kitt Peak synoptic maps of longitudinal
magnetic field. The magnetic butterfly diagram is continuously
constructed from synoptic magnetic maps measured at National Solar
Observatory by averaging the magnetic flux in longitude at each
latitude for each solar rotation since 1976 \citep{hat03}.

\begin{figure}
\epsscale{0.45} \plotone{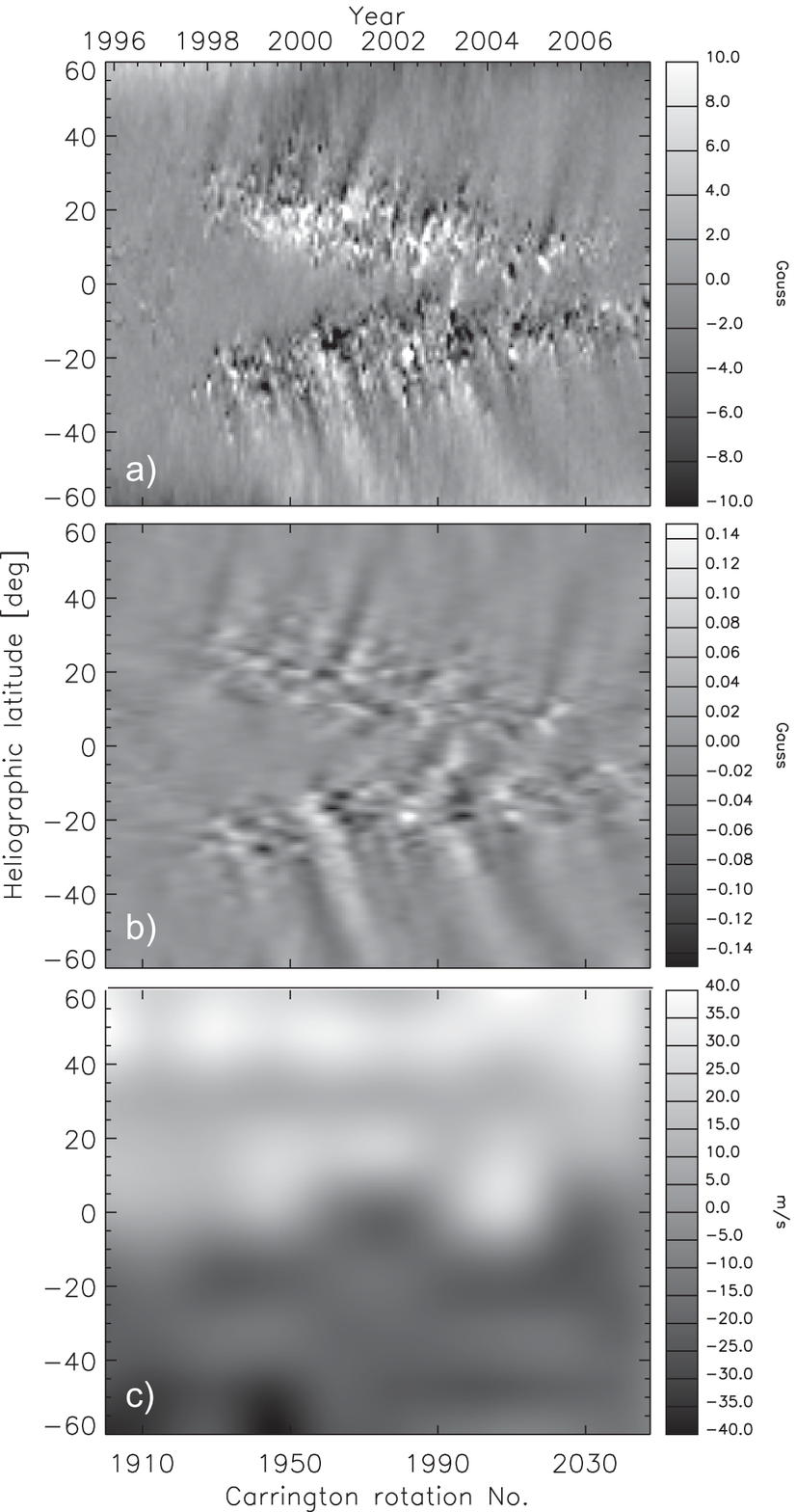} \caption{a) The magnetic butterfly
diagram for cycle 23. b) The filtered magnetic butterfly diagram
showing enhancements of the flux transport elongated structures. c)
The measured meridional flux transportation speed (in the
South--North direction)  for Carrington rotations
1900--2048.\label{maps}}
\end{figure}

At mid-latitudes of the magnetic butterfly diagram (Fig.~\ref{maps}a), between the active region zone and polar regions,
we clearly see elongated structures corresponding to the
poleward magnetic flux transport. The aim of our method is to measure the slopes of these structures
and to derive the speed of the meridional magnetic flux transport.

In addition to the large-scale structures, the original diagram
contains small-scale relatively short-lived local
magnetic field structures, which appear as a `noise' in the diagram.
To improve the signal-to-noise ratio for
the magnetic flux structures we applied a frequency band-pass filter for the
frequencies between $1.06\times 10^{-8}$~s$^{-1}$ (period of
1093~days) and $3.17\times 10^{-7}$~s$^{-1}$ (period of 36.5 days).
The filtering procedure is performed separately for each individual
latitudinal cut on the diagram. We tried also other methods of enhancement of structures
and found that they all provided comparable results. The difference between  the original butterfly diagram and the filtered
one can be seen in Fig.~\ref{maps} (\emph{a} and \emph{b}). The flux-transport elongated structures
are more easily visible after the filtering, and therefore more
suitable for analysis.

The meridional flux transport speed is measured on the basis of 
cross-correlation of two rows, which are assumed to be similar in shape, 
but positions of structures are different due to
their meridional transport. We cross-correlate pairs of rows separated
by heliographic latitude $\Delta b$ in a sliding
window with the size of 55 Carrington rotations. The edges of
the correlation window are apodized by a smooth function to avoid
the boxcar effects. The extremal position is calculated as a maximum
of the parabolic fit of the set of correlation coefficients of
correlated windows in five discrete displacements. If the
distribution of the correlation coefficients does not have a maximum,
or if the $R^2$ is too low (under 0.8), the meridional velocity in this
pixel is not evaluated. We have chosen $\Delta b=5^\circ$ as the
best tradeoff between the spatial resolution and precision.

To make the procedure more robust, we average the calculated
meridional velocity for five consecutive frames separated by
0.5\degr{} and centered at $\Delta b$ from the studied row. If any
of the speeds in averaged five rows is far out of the expected range
($-$60 to $+$60 m\,s$^{-1}$), then it is not used in the averaging.
From the fit, the accuracy of the measured flow speed is evaluated
and the maximum value of the set of five independent measurements at
different rows is taken. The same procedure is done with the processed map rotated by
180\,$^\circ$ to avoid any possible preferences in the direction
determination, and both results are averaged. The measured errors
were taken as the maximum value of both independent measurements.

For the ongoing analysis, only the speeds that were measured with
the error lower than 3~m\,s$^{-1}$ were taken into account. This criterion and some
failures of the slope measurement introduce gaps in the data, which
we need to fill. For this purpose we need to determine the best
continuous differentiable field that approximates the data. The
determination of such a field can be done in various ways, but we
wish to avoid possible artifacts. For filling the gaps we used
the MultiResolution Analysis. It is based on wavelet analysis, and we have chosen the
Daubechies wavelet due to its compact support. This property is
important since it minimizes edge effects. Moreover, using
these wavelets also preserves the location of zero-crossing and
maxima of the signal during the analysis. Daubechies are claimed to
be very stable in the noisy environments. For details see
\cite{rieutord07}. 

The reconstructed meridional flux transport speed map can be seen in 
Fig.~\ref{maps}\emph{c}. The flux transport remains poleward during the whole
studied period. 

\begin{figure}
 \epsscale{0.8}
\plotone{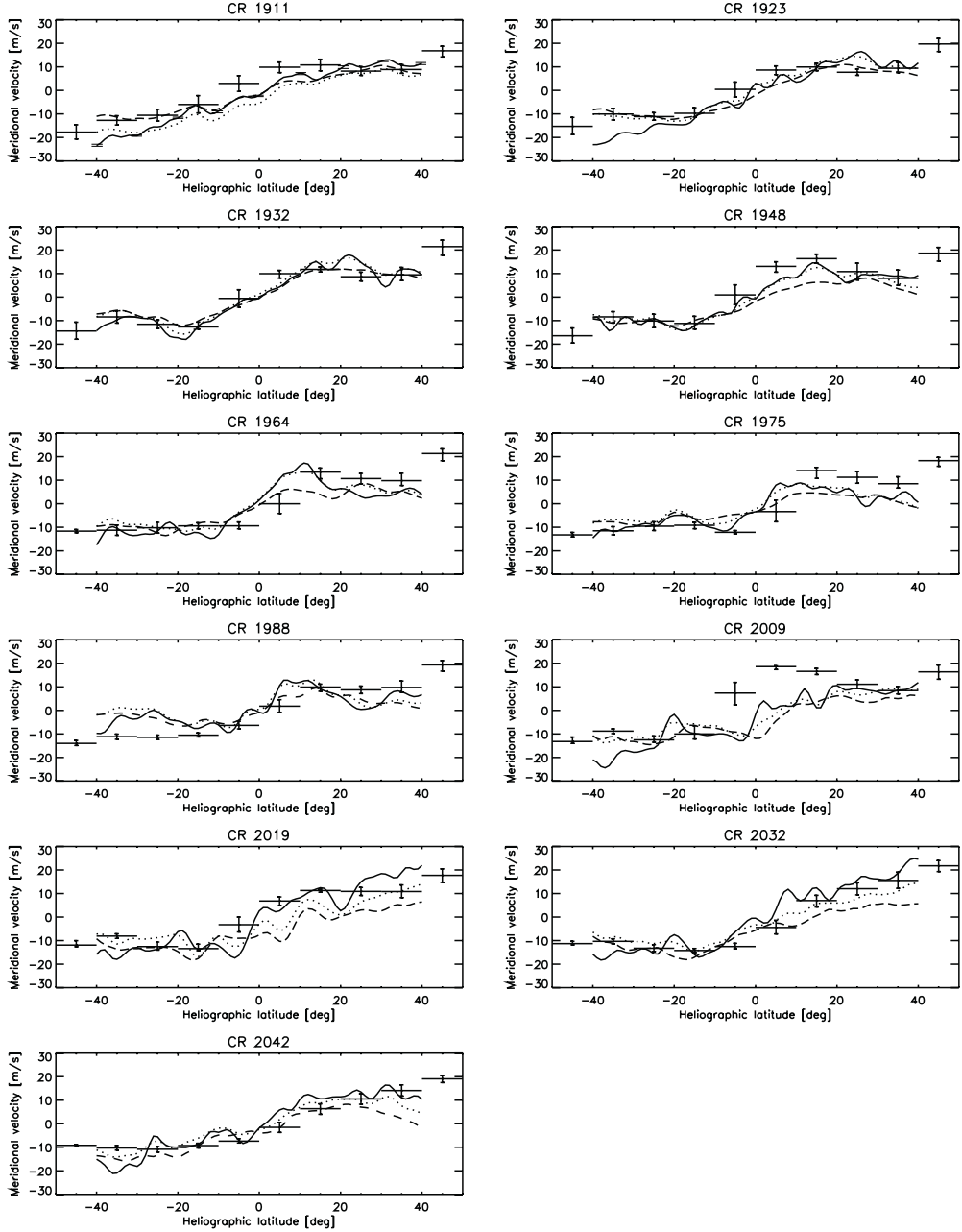} \caption{The longitudinally averaged meridional
flow speed measured for a set of Carrington rotations by
time-distance helioseismology. Solid line plots the time-distance
mean meridional flow at 3--4.5~Mm depth, dotted at 6--9~Mm, and
dashed at 9--12~Mm. The dots with error-bars represent the
10-degree-bin-averaged values of the flux transport speed derived
from the magnetic butterfly diagram (Fig~\ref{maps}a). Error bars of 
time-distance measurements are included for reference in CR~1911 
\citep[for details, see][]{2004ApJ...603..776Z}.
\label{rotations}}
\end{figure}

For the comparison between the meridional flow obtained from
time-distance helioseismology \citep{2004ApJ...603..776Z} and the
magnetic flux transport from our method, we have calculated the
averaged values of both quantities in bins of 10 heliographic degrees.
For 1996--2006, only eleven Carrington rotations have been evaluated
(one per year) by time-distance helioseismology using the full-disc Dynamics
data from the MDI instrument on SoHO spacecraft. These data are available
only for approximately two months per year. We compared the measurements of the magnetic flux
transport and the meridional flows in those
particular  non-consecutive Carrington rotations. The plots are
displayed in Fig.~\ref{rotations}.

\section{Results}

The magnetic flux transport speed and the mean meridional flow speed obtained from helioseismology 
are very similar (correlation coefficients are in
a range of 0.7--0.9). We have to keep in mind that while the time-distance meridional circulation profiles
represent the behaviour of the plasma during particular Carrington rotations, the magnetic butterfly diagram tracking
profiles represent the flux transport smoothed over 10 Carrington rotations. Therefore the agreement
cannot be perfect in principle. To make our results more accurate, the continuous helioseismic data are needed.

The speeds of the meridional flux transport in the near-equatorial region are less reliable, since the elongated structures in the magnetic butterfly diagram extend from the activity belts towards poles. In the equatorial region, significant parts of the measurements were excluded from the analysis due to their large measured error. The gaps were filled using MultiResolution Analysis from well-measured points. Although the results seem reasonable here, their lower reliability has to be kept in mind. The original data are constructed from the images obtained with low resolution with the orthographical projection to disc. Therefore values above the latitude of 50\,$^\circ$ are impacted by the projection effect, which reduces the spatial resolution and may cause an apparent increase of the measured meridional flux transport speed. This effect should be reduced if the higher-resolution data would be used. 

During the minimum of solar activity (such as CR 1911, 1923, or 2032), the profile of the meridional flux transport speed is very consistent with the mean longitudinally averaged profile of the meridional flow from helioseismology. The best agreement is found for depth of 9--12~Mm. This suggests that the flux transport may be influenced by flows in the deeper layers.

With increasing magnetic activity in the photosphere of the Sun, the gradient of the mean meridional circulation profile derived from time-distance helioseismology becomes steeper. This is consistent with the results obtained
by numerical simulations by \cite{2004SoPh..220..333B}. The simulations show that with increasing magnetic activity, the
Maxwell stresses oppose the Reynolds stresses, causing an acceleration of the meridional circulation and deceleration of
the rotation in low latitudes. Our measurements show that the variations of the slope of the mean meridional flux transport speed in latitude are lower with the progression of the solar cycle.

\begin{figure}
\epsscale{1} \plotone{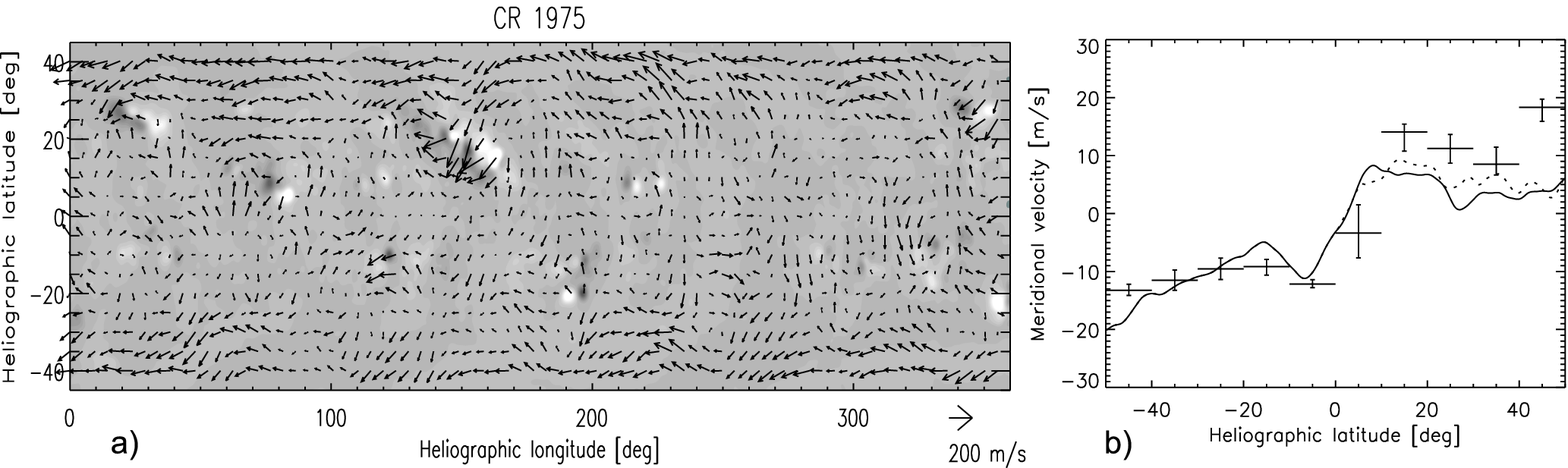} \caption{a) A large-scale flow map at
depth 3--4.5~Mm for Carrington rotation No. 1975 (April 2001) and the
corresponding MDI magnetogram in the grey-scale background.
Large-scale flows towards the equator in the magnetic regions are
visible around the large active region. b) The longitudinally
averaged meridional circulation profile for the same Carrington
rotation. The southern hemisphere depicts almost no magnetic
activity, so the meridional circulation profile obtained by
averaging the time-distance flow map (solid line) almost fits the
magnetic flux transport profile (points with error-bars) there,
while on the northern hemisphere they differ. After masking the
magnetic regions on the northern hemisphere, the recalculated
profile (dotted line) tends to fit the butterfly tracking one also
on the northern hemisphere. \label{casestudy}}
\end{figure}

When the large-active regions emerge in the activity belt, the flow towards equator is formed on the equatorial side of the magnetic regions (see example of the subsurface flow map in Fig.~\ref{casestudy}\emph{a}). This equatorward flow acts as a counter-cell of the meridional flow (present at the same longitudes as the corresponding magnetic region) and causes a decrease of the mean meridional flow amplitude in the activity belt. This behaviour is noticed in all studied cases recorded during eleven non-consecutive solar rotations, for which the Dynamics data useful for helioseismic inversion exist. Therefore the formation of the apparent counter-cell seems to be a common property of all large active regions in depths 3--12~Mm.

Flows in this counter-cell do not influence the magnetic flux transport, which can be demonstrated when the magnetic region is excluded from the synoptic map (Fig.~\ref{casestudy}\emph{b}). The calculated meridional circulation profile is then closer to the profile of the meridional flux transport speed derived from the magnetic butterfly diagram.

\section{Conclusions}

We have compared the measurements of the meridional speed derived from two
different techniques: by time-distance local helioseismology and by
measuring the flux transport speed using the magnetic butterfly
diagram. We have found that the results agree quite well in
general, but they differ in regions occupied by local
magnetic fields. The detailed flow maps from helioseismology show that this is partly due to the presence of meridional counter-cells at the
equatorial side of magnetic regions, which influence the time-distance derived meridional
flow profile, but does not influence the magnetic flux transport.

We have studied eleven non-consecutive Carrington rotations covering one solar cycle. The effect of the
local flows around active regions and especially on their equatorial side is noticed in all the studied
cases. Therefore, this behaviour seems to be a common property of the subsurface dynamics around active regions
located in the activity belt. However, we have to keep in mind that
both datasets are not directly comparable, since the time-distance flow maps represent the behaviour of
flows during one Carrington rotation, while the butterfly diagram tracking procedure provide results
averaged over few Carrington rotations. The more homogeneous data for local helioseismology are needed
to study this effect in more detail.

The results show that the speed of the magnetic flux transport
towards the solar poles may significantly deviate from the
longitudinally averaged meridional flow speed derived from local
helioseismology measurements, which are affected by local
circulation flows around active regions in the activity belt.
Therefore, using the longitudinally averaged meridional flow profile
from helioseismology in the solar cycle models for description the
flux transport is not justified. The longitudinal structure of these
flows should be taken into account.

\acknowledgments \sloppypar M\v{S} would like to gratefully acknowledge
ESA-PECS for the support under grant No.~8030, the Ministry of
Education of the Czech Republic under Research Program
MSM0021620860, and Stanford Solar Physics Group for support and hospitality.

\clearpage

\clearpage

\end{document}